# Joint user association and power allocation in ultra-dense mmWave networks: a multi-connectivity approach

Ailing Chen, Shengchang Li, Jichen Xiong, Kezhong Jin, and Zhenzhou Tang, *Senior Member, IEEE*

*Abstract*—In ultra-dense millimeter wave (mmWave) networks, mmWave signals suffer from severe path losses and are easily blocked by obstacles. Meanwhile, ultra-dense deployment causes excessive handovers, which reduces the data link reliability. To alleviate the above issues, the novel technology, known as multi-connectivity enabled user association (MCUA) is incorporated in this letter. We aim to jointly optimize MCUAs and downlink (DL) power allocations (PAs) to maximize the DL rate of each user simultaneously, rather than total. This is a non-convex nonlinear 0-1 mixed integer multi-objective optimization problem and quite complicated. To solve it, we first use the weighted sum method to scalarize it as a single-objective optimization problem (SOOP), and then relax the binary association variables to real ones. Considering that the relaxed SOOP is still non-convex, we perform a series of transformations upon it and make it a differential of convex programming. Finally, we develop an iterative algorithm based on the convex-concave procedure to solve the SOOP. Numerical results are presented to demonstrate the effectiveness of the proposed algorithms.

## I. Introduction

MILLIMETER wave (mmWave) has been an essential part of 5G and beyond 5G heterogeneous networks. However, there are still some issues that need to be addressed. For example, due to the ultra-high operating frequency, the path loss of mmWave signals is extremely serious [1]. Moreover, mmWave signals are easily blocked by obstacles [2]. Given this, in actual scenarios, mmWave networks need to be ultra-densely deployed to provide considerable transmission efficiency [3]. In addition, existing work has also shown that the use of multi-connectivity (MC) technology, which allows a user equipment (UE) to be associated with multiple mmWave base stations (mBSs), can effectively improve the reliability of data connections [4]. Therefore, a new problem naturally arises, that is, how to best associate UEs with mBSs and allocate DL power for each UE-mBS pair to achieve the best performance.

A. Chen, S. Li, J. Xiong, K. Jin and Z. Tang are with the College of Computer Science & Artificial Intelligence, Wenzhou University, Wenzhou, China, 325035.
K. Jin and Z. Tang are with the Innovation Research Center for Intelligent Networking, Wenzhou University, China, 325035.
This work was supported by the Zhejiang Provincial Natural Science Foundation of China under Grant LZ20F010008.
A. Chen and S. Li contributed equally overall to this work.
Corresponding author: Zhenzhou Tang (mr.tangzz@gmail.com).

Recently, the issue of joint optimization of multi-connectivity enabled user association (MCUA) and resource allocation in ultra-dense mmWave networks has been undergoing a heated discussion. In [5], the authors proposed a heuristic algorithm to solve the optimal MCUA problem to minimize the total power consumption optimization of 5G Heterogeneous Cloud Radio Access Network (HCRAN). In [6], a multi-connectivity enabled uplink measurement system was implemented to continuously monitor the channel propagation conditions in mmWave networks. Liu et al. developed a multi-label classification method to optimally associate UEs to multiple mBSs in mmWave networks with the objective of maximizing the system throughput [7]. In [8], a heuristic solution was proposed to solve the DL MCUA problem to improve the total throughput of small cellular networks. In [9], an optimal MCUA-PA method based on the constrained non-dominated sorting genetic algorithm II was developed to simultaneously maximize the system energy efficiency, balance the user achievable rates and balance the traffic loads of mSBs. The work in [10] proposed an iterative gradient algorithm to jointly optimize user association (UA) and power allocation.

However, the above work only optimizes the overall system performance. Considering that different users generally have different quality of service (QoS) requirements, the QoS for each user should be considered individually, rather than together. However, the QoS of different users cannot be maximized simultaneously since the QoS of each user is mutually restricted in resource-constrained systems. Given this, the tradeoff among the QoS of different users is more desirable in practical scenarios.

Motivated by the above, in this letter, we study the optimal MCUAs and DL PAs in ultra-dense mmWave networks to jointly maximize the DL rate of each user under the premise of ensuring QoS. It is a constrained nonlinear mixed 0-1 integer multi-objective optimization problem (MOOP) that not only optimizes the MCUA of each user, but also optimizes the DL PA of each user-mBS pair. Moreover, this MOOP is non-convex and NP-hard.

In order to solve this so complicated MOOP, we first relax the binary association variables to real ones in the range of $[0, 1]$. Then, the weighted sum method is used to scalarize the MOOP into a single objective optimization problem (SOOP). However, the SOOP is still a non-convex optimization problem. Therefore, we further transform the SOOP into an equivalent difference of convex (d.c.)

programming by introducing auxiliary variables, and finally solve it by developing an iterative algorithm based on the convex-concave procedure (CCP) [11].

The rest of this letter is organized as follows. The system model is shown in Section II. Then, we present the CCP based MCUA-PA method in Section III. Afterwards, the simulation results are shown and analyzed in Section IV. Finally, Section V concludes the work.

## II. System models

We consider an ultra-dense multi-connectivity enabled mmWave network. Denote the set of mBSs and the set of users as $\mathbb{B} = \{mBS_j, j \in \mathbb{I}_b = \{1, 2, \ldots, m\}$ and $\mathbb{U} = \{u_i\}, i \in \mathbb{I}_u = \{1, 2, \ldots, n\}$, respectively, where $m$ is the number of mBSs and $n$ is the number of users. Assuming a Rayleigh fading channel with a zero mean and unit variance and that $u_i$ is associated to $mBS_j$, the DL received power at $u_i$ can be written as

$$P_{ij}^r = P_{ij}^t d_{ij}^{-\alpha} h_{ij} G_a \left(\frac{\lambda}{4\pi}\right)^2, \quad (1)$$

where $P_{ij}^t$ is the DL transmission power from $mBS_j$ to $u_i$, $d_{ij}$ is the distance between $u_i$ and $mBS_j$, $\alpha$ is the path loss exponent, $h_{ij}$ is the channel coefficient, $\lambda$ is the wavelength, and $G_a$ is the gain of the beam from $mBS_j$ to $u_i$ and we assume that

$$G_a = \begin{cases} 1, \text{if } u_i \text{ locates within the beam of } mBS_j, \\ 0, \text{otherwise}, \end{cases} \quad (2)$$

To simplify the analysis, we assume that the direction of the beam always matches the connecting lines from $mBS_j$ to $u_i$. In other words, $G_a = 1$ if $u_i$ is associated to $mBS_j$.

In this study, we use the signal-to-noise-ratio (SNR) instead of the signal-to-interference-plus-noise-ratio (SINR) at the receiver, since mmWave networks are noise rather than interference-limited due to the highly directional beamforming and sensitivity to blockage [2], [12]. The assumption that large bandwidth mmWave networks are noise-limited has been considered and motivated in [12]. In addition, the study in [2] has validated (in Section V.B.2) that this assumption holds even for high densities of mmWave networks. Henceforth, this assumption is considered for this study. The SNR at $u_i$ due to the transmission of $mBS_j$ can be calculated as $\gamma_{ij} = \frac{P_{ij}^r}{W\sigma^2}$ where $W$ is the bandwidth and $\sigma^2$ is the power density of additive white Gaussian noise (AWGN).

So far, the achievable DL rate from $mBS_j$ to the $u_i$ can be obtained as

$$R_{ij} = \frac{1}{\sum_{i=1}^n x_{ij}} W x_{ij} \log_2\left(1 + \gamma_{ij}\right), \quad (3)$$

where $x_{ij} \in \{0, 1\}$ is the association indicator between $u_i$ and $mBS_j$ and

$$x_{ij} = \begin{cases} 1, \text{if } u_i \text{ is associated with } mBS_j, \\ 0, \text{otherwise}, \end{cases} \quad (4)$$

$\sum_{i=1}^n x_{ij}$ is the number of users associated with $mBS_j$. And the total DL rate at $u_i$ is $R_i = \sum_{j \in \mathbb{I}_b} R_{ij}$.

In this letter, we aim to jointly maximize the total achievable rate of each user in the multi-connectivity enabled mmWave network and achieve the QoS balance among all users on the premise of ensuring the given QoS constraints through optimally associating each user to one or more mBSs and allocating DL transmission power for each mBS-user pair. Consequently, the constrained optimization problem can be formulated as

$$OP1: \max_{\mathbf{X}, \mathbf{P}} \{R_1, R_2, \ldots, R_n\}, \quad (5)$$

subject to

$$C1: x_{ij} \in \{0, 1\}, \forall i \in \mathbb{I}_u, \forall j \in \mathbb{I}_b,$$
$$C2: \sum_{j \in \mathbb{I}_b} x_{ij} \geq 1, \forall i \in \mathbb{I}_u,$$
$$C3: \sum_{i \in \mathbb{I}_u} x_{ij} \geq 1, \forall j \in \mathbb{I}_b,$$
$$C4: P_{\min} \leq P_{ij}^t \leq P_{\max}, \forall i \in \mathbb{I}_u, \forall j \in \mathbb{I}_b,$$
$$C5: \sum_{i \in \mathbb{I}_u} x_{ij} P_{ij}^t \leq P_{\max}, \forall j \in \mathbb{I}_b,$$
$$C6: R_i \geq R_{\min}, \forall i \in \mathbb{I}_u,$$

where $\mathbf{X} = \{\mathbf{x}_j, j \in \mathbb{I}_b\}$ is the user association vector, $\mathbf{x}_j = \{x_{ij}, i \in \mathbb{I}_u\}$, $\mathbf{P} = \{\mathbf{p}_j, j \in \mathbb{I}_b\}$ is the transmission power vector, $\mathbf{p}_j = \{P_{ij}^t, i \in \mathbb{I}_u\}$, $R_{\min}$ is the minimum achievable rate requirement for $u_i$, and $P_{\min}$ and $P_{\max}$ are the minimum and maximum transmission power of an mBS, respectively. The constraint $C2$ and $C3$ stipulate that each user must be attached to at least one mBS and each mBS should associate with at least one user, respectively, $C4$ limits the maximum total transmission power of each mBS, $C5$ presents the range of the DL transmission power, and $C6$ ensures the QoS constraints.

**Proposition 1.** *The optimization problem formulated in* (5) *is an NP-hard problem.*

*Proof:* The optimization problem formulated in (5) can be reduced to a multi-knapsack problem (MKP) which has been proved to be NP-complete in polynomial time. ∎

## III. Methodology

To solve the MOOP formulated in (5), firstly, we use the traditional weighted sum method to scalarize the MOOP into a single objective problem. However, it is still a mixed integer programming and non-convex optimization problem. In view of this, we relax the binary association variables to real ones in the range of [0, 1]. After that, by introducing relaxation variables, it can be transformed into an equivalent d.c. programming problem, which is then solved by developing an iterative algorithm based on the CCP method.

### A. Scalarization & Relaxation

The weighted sum method, which is one of the most common methods for solving multi-objective optimization, is used to combine all the objectives together to form a single objective function. For this purpose, the problem OP1 in (5) can be rewritten as

$$OP2: \max_{\mathbf{X}, \mathbf{P}} \left\{\sum_{i \in \mathbb{I}_u} w_i R_i\right\}, \quad (6)$$

subject to $C1 \sim C6$, where $w_i$ is the weight for the $i$th objective ($R_i$) and $\sum_{i=1}^n w_i = 1$.

Considering that $x_{ij}$ defined in OP2 is a binary integer, we relax $x_{ij}$ to a real number $\tilde{x}_{ij} \in [0, 1]$ to facilitate solving OP2 as follows:

$$\text{OP2r}: \max_{\tilde{\mathbf{X}},\mathbf{P}} \sum_{i \in \mathbb{I}_u} w_i \tilde{R}_i, \quad (7)$$

subject to $C4$ and

$$\tilde{C}1: \tilde{x}_{ij} \in [0, 1], \forall i \in \mathbb{I}_u, \forall j \in \mathbb{I}_b,$$

$$\tilde{C}2: \sum_{j \in \mathbb{I}_b} \tilde{x}_{ij} \geq 1, \forall i \in \mathbb{I}_u,$$

$$\tilde{C}3: \sum_{i \in \mathbb{I}_u} \tilde{x}_{ij} \geq 1, \forall j \in \mathbb{I}_b,$$

$$\tilde{C}5: \sum_{i \in \mathbb{I}_u} \tilde{x}_{ij} P_{ij}^t \leq P_{\max}, \forall j \in \mathbb{I}_b,$$

$$\tilde{C}6: \tilde{R}_i \geq R_{\min}, \forall i \in \mathbb{I}_u,$$

where $\tilde{\mathbf{X}} = \{\tilde{x}_{ij}, i \in \mathbb{I}_u, j \in \mathbb{I}_b\}$ and

$$\tilde{R}_{ij} = \frac{1}{\sum_{i=1}^n \tilde{x}_{ij}} W \tilde{x}_{ij} \log_2\left(1 + \gamma_{ij}\right). \quad (8)$$

By introducing two auxiliary variable vectors $\mathbf{u} = \{u_{ij}\}$ and $\mathbf{v} = \{v_{ij}\}$, OP2r can be further transformed to the following optimization problem:

$$\text{OP3}: \max_{\tilde{\mathbf{X}},\mathbf{P},\mathbf{u},\mathbf{v}} \sum_{i \in \mathbb{I}_u} \sum_{j \in \mathbb{I}_b} u_{ij}, \quad (9)$$

subject to $\tilde{C}1, \tilde{C}2, \tilde{C}3, C4, \tilde{C}5$ and

$$C7: \sum_{j \in \mathbb{I}_b} u_{ij} \geq w_i R_{\min}, \forall i \in \mathbb{I}_u, \quad (10d)$$

$$C8: \frac{v_{ij}^2}{\sum_{\mathbb{I}_u} \tilde{x}_{ij}} \geq u_{ij}, \forall i \in \mathbb{I}_u, \forall j \in \mathbb{I}_b, \quad (10i)$$

$$C9: W \tilde{x}_{ij} w_i \log_2\left(1 + \gamma_{ij}\right) \geq v_{ij}^2, \forall i \in \mathbb{I}_u, \forall j \in \mathbb{I}_b. \quad (10h)$$

Apparently, the constraints $C8$ and $C9$ hold with equality at optimality, so problem OP3 and OP2r are equivalent.

### B. Solution of OP3

Firstly, $\tilde{R}_i, \forall i \in \mathbb{I}_b$ in OP2r is a fraction and non-convex. To facilitate solving it, we have the following theorem.

**Theorem 1.** *OP3 can be iteratively solved by the following convex programming:*

$$\text{OP4}: \max_{\mathbf{X},\mathbf{P},\mathbf{u},\mathbf{v}} \sum_{i \in \mathbb{I}_u} \sum_{j \in \mathbb{I}_b} u_{ij}, \quad (10)$$

subject to $\tilde{C}1, \tilde{C}2, \tilde{C}3, C4, C7$ and

$$C10: \frac{1}{4} \sum_{i \in \mathbb{I}_u} \left(\left(\tilde{x}_{ij} + P_{ij}^t\right)^2 - f\left(\tilde{x}_{ij}, P_{ij}^t\right)\right) \leq P_{\max}, \forall j \in \mathbb{I}_b,$$

$$C11: \frac{2\dot{v}_{ij} v_{ij}}{\sum_{\mathbb{I}_u} \dot{x}_{ij}} - \frac{\dot{v}_{ij}^2 \sum_{\mathbb{I}_u} \tilde{x}_{ij}}{\left(\sum_{\mathbb{I}_u} \dot{x}_{ij}\right)^2} \geq u_{ij}, \forall i \in \mathbb{I}_u, \forall j \in \mathbb{I}_b,$$

$$C12: W w_i \log_2\left(1 + \gamma_{ij}\right) \geq v_{ij}^2, \forall i \in \mathbb{I}_u, \forall j \in \mathbb{I}_b,$$

*where the given points $\left(\dot{x}_{ij}, \dot{P}_{ij}^t, \dot{u}_{ij}, \dot{v}_{ij}\right), (\forall i \in \mathbb{I}_u, \forall j \in \mathbb{I}_b)$ in the $k$th iteration can be obtained as*

$$\left(\dot{x}_{ij}, \dot{P}_{ij}^t, \dot{u}_{ij}, \dot{v}_{ij}\right)\bigg|k = \arg\max_{\mathbf{X},\mathbf{P},\mathbf{u},\mathbf{v}} \sum_{i \in \mathbb{I}_u} \sum_{j \in \mathbb{I}_b} u_{ij}\bigg|k - 1, \quad (11)$$

*subject to $\tilde{C}1, \tilde{C}2, \tilde{C}3, C4, C7, C10, C11$ and $C12$.*

*Proof:* First of all, we prove that OP3 can be rewritten to a d.c. programming, that is OP4.

It is straightforward that the objective, $\tilde{C}1, \tilde{C}2, \tilde{C}3, C4$ and $C7$ in OP3 are linear.

The total transmission power of $\text{mBS}_j$ is constrained by $\tilde{C}5$ in OP2r, where

$$\tilde{x}_{ij} P_{ij}^t = \frac{1}{4}\left(\tilde{x}_{ij} + P_{ij}^t\right)^2 - \frac{1}{4}\left(\tilde{x}_{ij} - P_{ij}^t\right)^2. \quad (12)$$

Apparently, it is a standard d.c. form since both the first and second terms on the right-hand side of the equation are convex.

In $C8$, $u_{ij}$ is linear and semi-convex. Besides, $\frac{v_{ij}^2}{\sum_{\mathbb{I}_u} \tilde{x}_{ij}}$ is a quadratic fraction function which is also convex. So $C8$ is a d.c. constraint.

Consequently, OP4 is a d.c. programming and can be solved by CCP [11].

Next, we convexify $\tilde{C}5, C8$ and $C9$ of OP3 as follows.

The lower bound of $\frac{1}{4}\left(\tilde{x}_{ij} - P_{ij}^t\right)^2$ can be obtained by taking its first-order Taylor series expansion at the point $\left(\dot{x}_{ij}, \dot{P}_{ij}^t\right)$. So we have the following convex constraint:

$$C10: \frac{1}{4} \sum_{i \in \mathbb{I}_u} \left(\left(\tilde{x}_{ij} + P_{ij}^t\right)^2 - f\left(\tilde{x}_{ij}, P_{ij}^t\right)\right) \leq P_{\max}, \forall j \in \mathbb{I}_b, \quad (13)$$

where

$$f\left(\tilde{x}_{ij}, P_{ij}^t\right) = \left(\dot{x}_{ij} - \dot{P}_{ij}^t\right)^2 + 2\left(\dot{x}_{ij} - \dot{P}_{ij}^t\right)\left(\tilde{x}_{ij} - \dot{x}_{ij} - P_{ij}^t + \dot{P}_{ij}^t\right).$$

For the fractional constraint $C8$, by taking the first-order Taylor series expansion of $\frac{v_{ij}^2}{\sum_{\mathbb{I}_u} \tilde{x}_{ij}}$ at $\left(\dot{x}_{ij}, \dot{v}_{ij}\right)$, we obtain the $C11$ in OP4:

$$C11: u_{ij} - \left(\frac{2\dot{v}_{ij} v_{ij}}{\sum_{\mathbb{I}_u} \dot{x}_{ij}} - \frac{\dot{v}_{ij}^2 \sum_{\mathbb{I}_u} \tilde{x}_{ij}}{\left(\sum_{\mathbb{I}_u} \dot{x}_{ij}\right)^2}\right) \leq 0. \quad (14)$$

For logarithmic constraint $C9$, we have $W w_i \log_2\left(1 + \gamma_{ij}\right) \geq W w_i \tilde{x}_{ij} \log_2\left(1 + \gamma_{ij}\right)$, since $\tilde{x}_{ij} \in [0, 1]$. So we have

$$C12: W w_i \log_2\left(1 + \gamma_{ij}\right) \geq v_{ij}^2, \forall i \in \mathbb{I}_u, \forall j \in \mathbb{I}_b. \quad (15)$$

Finally, OP4 can be solved iteratively by CCP, as shown in Algorithm 1. The detailed convergence proof of Algorithm 1 is provided in [11].

So far, we have solved OP2r which is the relaxation of OP2. And the integer solutions of OP2 can be easily obtained by branch-and-bound method or some other mixed integer programming methods. ∎

**Algorithm 1** Convex-concave procedure for solving OP4

1: Initialize $\tau > 0$, $k \leftarrow 1$, feasible $\mathbf{X}_0$, $\mathbf{P}_0^t$, $\mathbf{u}_0$ and $\mathbf{v}_0$.
2: $R_0 \leftarrow \sum_{i \in \mathbb{I}_u} \sum_{j \in \mathbb{I}_b} \mathbf{u}_0, \forall j \in \mathbf{I}_b$.
3: **repeat**
4: $\quad R_k \leftarrow \max_{\mathbf{X},\mathbf{P}^t,\mathbf{u},\mathbf{v}} \left\{ \sum_{i \in \mathbb{I}_u} \sum_{j \in \mathbb{I}_b} u_{ij} \right\}$, s.t. $\tilde{C}1 - \tilde{C}3, C4, C7, C10 - C12$.
5: $\quad k \leftarrow k + 1$
6: **until** $|R_k - R_{k-1}| \leq \tau$

TABLE I: Simulation Setup

| Parameter | Value |
|---|---|
| Bandwidth $W$ | 100 MHz |
| Path loss exponent $\alpha$ | 2 |
| Millimeter wave length $\lambda$ | 5 mm |
| Noise power spectral density $\sigma$ | -174 dBm/Hz |
| Minimum Rate $R_{\min}$ | 100 Mbit/s |
| Minimum/Maximum transmission power of an mBS | 0/1000 mW |

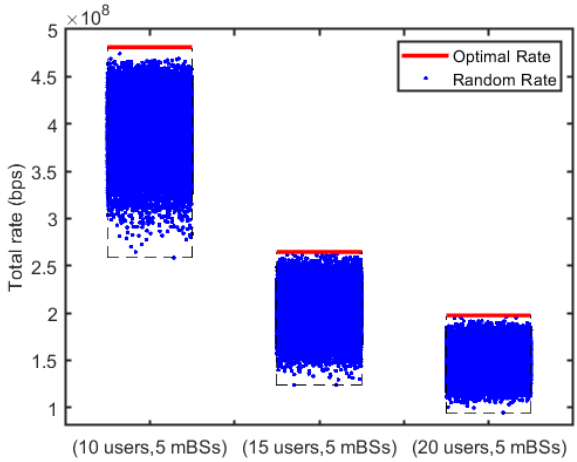

Fig. 1: Equal-weighted sum of the DL rate in different scenarios.

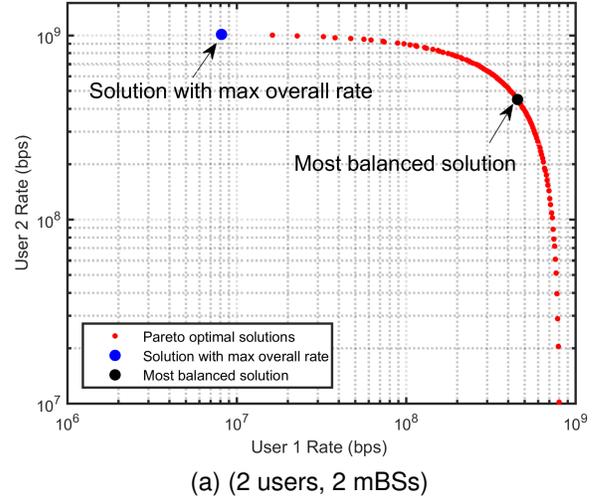

(a) (2 users, 2 mBSs)

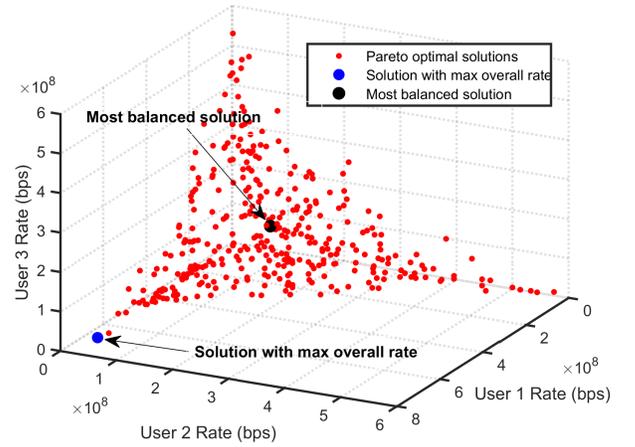

(b) (3 users, 2 mBSs)

Fig. 2: The optimal solutions obtained with different weight vectors.

## IV. Results and analysis

In this section, the effectiveness of the proposed MCUA-PA method is proved by a large number of experiments. We assume that the mBSs were fixed within the service area of 100 m × 100 m. Unless otherwise stated, the parameter settings of the simulations are shown in Table I. We use CVX [13] and MOSEK [14] to solve the convex problems in the proposed method.

We first conducted a Monte Carlo experiment to verify the effectiveness of the proposed MCUA-PA method. Specifically, we set up three scenarios of different scales, i.e., the (10 users, 5 mBSs) scenario, the (15 users, 5 mBSs) scenario, and the (20 users, 5 mBSs) scenario. In each scenario, the locations of mBSs and users were fixed. We randomly generated $10^5$ feasible MCUA-PA solutions in each scenario and the equal-weighted sum of the DL rate of each scenario was calculated. The results are shown in Fig. 1 where the blue dots represent the random solutions and the red line segments are the equal-weighted sum of the optimal solutions obtained by the MCUA-PA method, i.e., the optimal total DL rates. We can observe that the optimal total DL rates are always the highest, which proves that the proposed method is effective.

Fig. 2 shows the Pareto fronts (PFs) obtained by the proposed MCUA-PA method by varying the weight for each user between 0.01 to 0.99 with a step-size of 0.01, since each weight vector corresponding to a specific Pareto optimal MCUA-PA solution. Generally, the relative value of the weights reflects the relative importance of the objectives [15]. And the decision-makers can choose to generate the expected Pareto optimal MCUA-PA solutions with preset weight vectors. Fig. 2 marks out the most balanced solution and the solution with maximum total DL rate, respectively.

In order to investigate the effect of the maximum number of allowable associated mBSs upon the achievable DL rate, we also compared the optimal total DL rates obtained by the proposed MCUA-PA method and the PoP-UA-PA (UA-PA following the principle of proximity), respectively, under different maximum numbers of mBSs allowed for association. In PoP-UA-PA, a user is always associated with as many nearest mBSs as possible and the corresponding DL powers are optimally allocated. In this set of experiments, the locations of mBSs (five mBSs) were fixed and those of users formed an independent binomial point process. Each experiment was independently run 1000 times. The results are illustrated in Fig. 3. We can find that the optimal total DL rates are always greater than those obtained by PoP-UA-PA. In addition, the more mBS allowed to be associated, the greater the performance gap is.

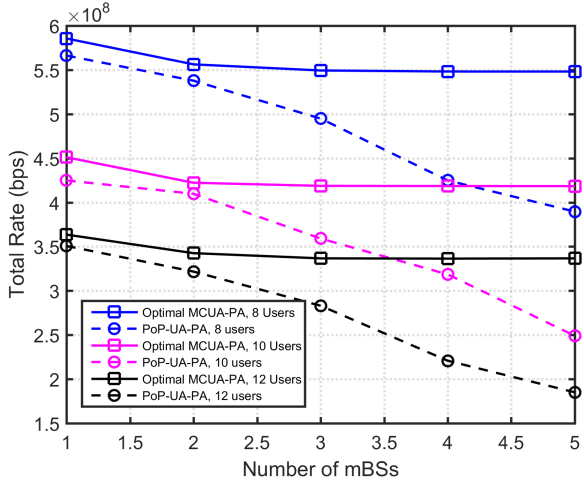

Fig. 3: Total DL rates under different maximum numbers of mBSs allowed to be associated, the proposed MCUA-PA method versus PoP-UA-PA.

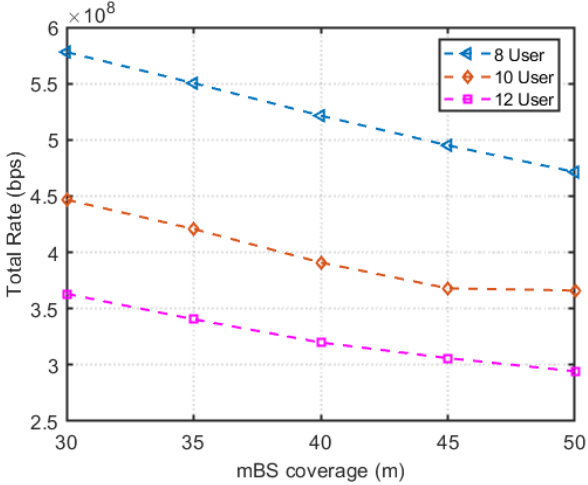

Fig. 4: Total rates under different mBSs coverages (5 mBSs).

Finally, we study how the mBS coverage affects the optimal DL rate. To this end, we compared the total DL rates under various mBS coverages with different number of users, as shown in Fig. 4. Also, five mBSs were involved in the experiments whose locations were fixed and the locations of users formed an independent binomial point process. Each experiment was independently run 1000 times. According to the results shown in Fig. 4, we can see that the optimal total DL rates decrease along with the increase of mBSs coverage. This is because a larger coverage may introduce users that are far away from the associated mBSs. And it is commonly known that the farther a user is from the mBS, the lower DL rate can be achieved.

## V. CONCLUSION

This letter proposes an iterative method based on CCP to find the optimal MCUA-PA to maximize the achievable DL rates of each user simultaneously under the quality of service (QoS) constraints. To solve this non-convex, NP-hard and 0-1 mixed integer MOOP, we first scalarize it to an SOOP by the weighted sum method. Then, we relax the 0-1 mixed integer SOOP and perform a series of problem transformations to convert it to a d.c. programming, which can be solved by the iterative CCP. Numerical results demonstrate the effectiveness of the proposed algorithms.